\documentclass[pre,aps,showpacs,twocolumn,floatfix]{revtex4}
\usepackage{graphicx}
\usepackage{dcolumn}
\usepackage{bm}

\newcommand{\be}{\begin{equation}}
\newcommand{\ee}{\end{equation}}
\newcommand{\bea}{\begin{eqnarray}}
\newcommand{\eea}{\end{eqnarray}}

\let\oldphi\phi
\let\phi\varphi
\let\varphi\oldphi

\newcommand{\dr}{\partial}
\newcommand{\nab}{\vec \nabla}

\newcommand{\ccc}{$\mbox{\textsf{C}}_{\mbox{\textsf{c}}}^{\mbox{\textsf{c}}}$}

\begin{document}
\title{Corridors of barchan dunes: stability and size selection.}
\author{
P. Hersen$^a$, K.H. Andersen$^b$, H. Elbelrhiti$^c$,
B. Andreotti$^a$, P. Claudin$^d$ and S. Douady$^a$
}
\affiliation{
$^a$
Laboratoire de Physique Statistique de l'ENS, 24 rue Lhomond,
75005 Paris, France.\\
$^b$
Department of Mechanical Engineering,
Technical University of Denmark,
DK-2800 Lyngby, Denmark. \\
$^c$
Universit\'e Ibn Zohr, Facult\'e des Sciences, BP 28/S,
Cit\'e Dakhla, 80000 Agadir, Morocco.\\
$^d$
Laboratoire des Milieux D\'esordonn\'es et H\'et\'erog\`enes (UMR 7603),
4 place Jussieu -- case 86, 75252 Paris Cedex 05, France.
}

\begin{abstract}
Barchans are crescentic dunes propagating on a solid ground.  They form
dune fields in the shape of elongated corridors in which the size and
spacing between dunes are rather well selected.  We show that even
very realistic models for solitary dunes do {\it not} reproduce these
corridors.  Instead, two instabilities take place.  First, barchans
receive a sand flux at their back proportional to their width while
the sand escapes only from their horns.  Large dunes proportionally
capture more than they loose sand, while the situation is reversed for
small ones: therefore, solitary dunes cannot remain in a steady state.
Second, the propagation speed of dunes decreases with the size of the
dune: this leads -- through the collision process -- to a coarsening
of barchan fields.  We show that these phenomena are not specific to
the model, but result from general and robust mechanisms.  The length
scales needed for these instabilities to develop are derived and
discussed.  They turn out to be much smaller than the dune field
length.  As a conclusion, there should exist further -- yet unknown --
mechanisms regulating and selecting the size of dunes.
\end{abstract}
\maketitle

Since the pioneering work of Bagnold \cite{B41}, sand dunes have 
become an object of research for physicists.  Basically, the 
morphogenesis and the dynamics of dunes result from the interaction 
between the wind, which transports sand grains and thus modifies the 
shape of the dune, and the dune topography which in turn controls the 
air flow.  A lot of works have been devoted to the study of the 
mechanisms at the scale of the grain 
\cite{O64,H77,S85,JS86,AH88,AH91,ASW91,S91,WMR91,RM91,NHB93,IR94,RIR96,RVB00,SKH01,A03} 
and at the scale of single dune 
\cite{B10,F59,C64,LS64,H67,N66,H87,S90,HH98,SRPH00,ACD02b,HDA02,M88,ZJ87,JH75,HMGP78,JZ85,WG86,HLR88,WHCWWLC91,W95,BAD99,KSH01,S01}. 
 The interested reader should refer to a previous paper \cite{ACD02a} 
for a review of these works.  Our aim is to focus here on dune fields 
and to show that most of the problems at this scale are still open or 
even ill-posed.
\begin{figure}[h!]
\vspace{-1mm} \includegraphics{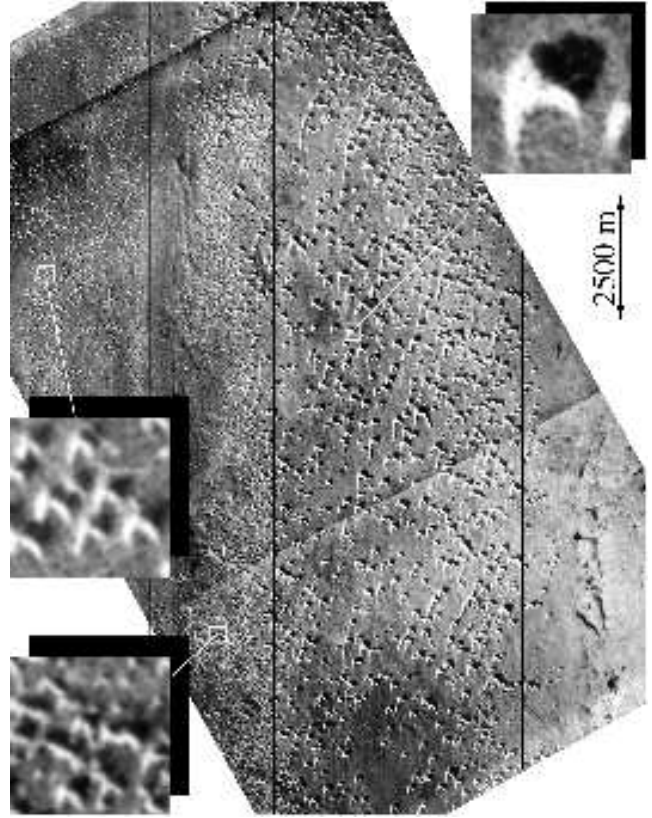} \vspace{-1mm} 
\caption{Aerial photograph showing part of the barchan field extending 
between Tarfaya, Laayoune and Sidi Aghfinir in southern Morocco, 
former Spanish Sahara.  The trade winds, dominant in the region, blow 
from the north (from the top of the photograph).  Several corridors 
are visible in which the size of barchans and their density is almost 
uniform.  As confirmed by the three zooms, the size of dunes is 
different from one corridor to another.}
\label{CoLayoune}
\vspace{-5mm}
\end{figure}

The most documented type of dune, the barchan \footnote{The success 
story of the barchan perhaps originates from the fact that cars can 
easily come close to their feet.}, is a crescentic shaped dune, horns 
downwind, propagating on a solid ground.  In the general picture 
emerging from the literature, barchans are thought as solitary waves 
propagating downwind without changing their shape and weakly coupled 
to their neighborhood.  For instance, most of the field observations 
concern geometric properties (morphologic relationships) and kinematic 
properties (propagation speed).  This essentially static description 
probably results from the fact that barchans do not change a lot at 
the timescale of one field mission.

As shown on figure \ref{CoLayoune}, barchans usually do not live 
isolated but belong to rather large fields \cite{LL69}.  Even though 
they do not form a regular pattern, it is obvious that the average 
spacing is a few times their size, and that they form long 
\emph{corridors} of quite uniformly sized dunes.  Observing the right 
part of figure~\ref{CoLayoune}, the barchans have almost all the same 
size ($6~m$ to $12~m$ high, $60~m$ to $120~m$ long and wide).  
Observing now the left part of figure \ref{CoLayoune}, the barchans 
are all much smaller ($1.5~m$ to $3~m$ high, $15~m$ to $30~m$ long and 
wide) and a small band can be distinguished, in which the density of 
dunes becomes very small.  Globally, five corridors stretched in the 
direction of the dominant wind can be distinguished: from right to 
left, no dune, large dunes, small dunes, no dune and small dunes 
again.  Figure~\ref{CoLayoune} shows only $17~km$ of the barchan 
field.  Direct observations show that these five corridors persist in 
a {\it coherent} manner over {\it hundreds} of kilometers along the 
dominant wind direction.

The content of this paper is perhaps a bit unusual as we will mostly 
present \emph{negative} results.  Indeed, we will show that none of 
the dune models 
\cite{HMGP78,JZ85,WG86,HLR88,WHCWWLC91,W95,BAD99,KSH01,S01}, nor the 
coarse grained field simulation \cite{LSHK02} are able at present to 
reproduce satisfactorily the selection of size and the formation of 
corridors.

More precisely, we shall first address the stability of solitary 
dunes, and conclude that, given reasonable orders of magnitude for 
dune sizes and velocities, barchans which are considered as 
``marginally unstable'' by other authors \cite{S01} would in fact 
have the time to develop their instability over a length much smaller 
than that of the corridor they belong to.  As a consequence, isolated 
dunes must be considered as truly unstable objects.  Furthermore, the 
origin of this instability is rather general and model independent, as 
it can be understood from the analysis of the output sand flux as a 
function of the dune size.

One can wonder whether interactions via collisions between dunes can 
modify the dynamics and the stability of dunes.  In a recent paper 
Lima \emph{et al.} \cite{LSHK02} have investigated the dynamics of a 
field and have claimed to get realistic barchan corridors.  However, 
they made use of numerical simulations into which individual dunes are 
stable objects of almost equal size ($6\%$ of
polydispersity).  They consequently obtained a nearly homogeneous field 
composed of dunes whose width is that of those injected at the upwind 
boundary. We show here that the actual case of individually unstable 
dunes leads by contrast to an efficient coarsening of the barchan 
field.

The paper is organized as follows.  In order to get a good idea of the 
mechanisms leading to these two instabilities, we first derive a 3D 
generalization of the \ccc\ model previously used to study 2D dunes 
\cite{ACD02b}.  We then show that the two instabilities predicted by 
the \ccc\ model are in fact very general and we will derive in a more 
general framework the time and length scales over which they develop.  
Turning to field observations, we will conclude that the formation of 
nearly uniform barchan corridors is an open problem: there should 
exist further mechanisms, not presently known and may be related to 
more complicated and unsteady effects such as storms or change of wind 
direction, to regulate the dune size.


\section{Barchan modeling. The \ccc\ model}
We start here with the state of the art concerning the modeling of 
dunes by Saint-Venant like equations.  First, the mechanisms of 
transport at the scale of the grain 
\cite{O64,S85,JS86,AH88,AH91,ASW91,S91,WMR91,RM91,NHB93,IR94,RIR96,RVB00,SKH01,A03,ACD02a} 
determine at the macroscopic scale -- at the scale of the dune -- the 
maximum quantity of sand that a wind of a given strength can 
transport.  As a matter of fact, when the wind blows over a flat sand 
bed, the sand flux increases and saturates to its maximum value $Q$ 
after a typical length $L$ called the saturation length 
\cite{B41,SKH01,ACD02b,HDA02}.  This length determines the size of the 
smallest propagative dune.

The other part of the problem is to compute the turbulent flow around 
a huge sand pile of arbitrary shape \cite{M88,ZJ87}.  Since the 
Navier-Stokes equations are far too complicated to be completely 
solved, people have derived simplified descriptions of the turbulent 
boundary layer 
\cite{JH75,HMGP78,JZ85,WG86,HLR88,WHCWWLC91,W95,BAD99,KSH01,S01}.  The 
first step initiated by Jackson, Hunt \emph{et al.} has been to derive 
an explicit expression of the basal shear stress in the limit of a 
very flat hill.  Kroy \emph{et al.} \cite{KSH01,S01} have shown that 
this expression can be simplified without loosing any important 
physical effect.  In particular, it keeps the non-local feature of the 
velocity field: the wind speed at a given place depends on the whole 
shape of the dune.

Being a linear expansion, this approach can not account for boundary
layer separation and in particular for the recirculation bubble that
occurs behind dunes.  Following Zeman and Jensen \cite{ZJ87} and later
Kroy \emph{et al.}, the Jackson and Hunt formula is in fact applied to
an envelope of the dune constituted by the dune profile prolonged by
the separation surface.

As already stated in one of our previous papers \cite{ACD02b}, we
proposed to name \ccc\ the class of models which describe the dynamics
of dunes in terms of the dune profile $h$ and the sand flux $q$, and
which include \emph{(i)} the mass conservation, \emph{(ii)} the
progressive saturation of sand transport and \emph{(iii)} the feedback
of the topography on the sand erosion/deposition processes.  We chose
this fancy name in reference to the spatial organization of the dunes
which propagate like the flight of wild ducks and geese.

\subsection{2D and 3D main equations}
Let us start with a quick recall of the set of 2D \ccc\ equations that
we already introduced in \cite{ACD02b}.  Let $x$ denote the axis
oriented along the wind direction, and $t$ the time.  The continuity
equation which ensures mass conservation simply reads
\be
\label{cc1}
\dr_t h + \dr_x q = 0.
\ee
Note that $q(x,t)$ denotes the integrated \emph{volumic} sand flux, 
i.e.  the volume of sand that crosses at time $t$ the position $x$ per 
unit time.  The saturation process is modeled by the following charge 
equation %
\be
\label{cc2}
\dr_x q = \frac{q_{sat} -q}{L}.
\ee
It is enough to incorporate the fact that the sand flux follows the
saturated flux $q_{sat}$ with a spatial lag $L$.  It is a linearized
version of the charge equation proposed by Sauermann {\emph{et al.}
\cite{SKH01}.

The saturated flux $q_{sat}$ is a growing function of the shear
stress.  This shear stress can be related to the dune profile $h$ by
the modified Jackson and Hunt expression.  Since this expression comes
from a linear expansion, we can directly relate $q_{sat}$ to $h$ by:
\be
\label{cc3}
\frac{q_{sat}(x)}{Q} = 1 + A \int \!\! \frac{d\chi}{\pi\chi} \, \dr_x
h_e(x-\chi) + B \, \dr_x h_e(x),
\ee
where $Q$ is the saturated flux on a flat bed and $h_e$ the envelope 
prolonging the dune on the lee side (see Appendix and 
\cite{KSH01,ACD02b} for the details of construction).  The last term 
takes into account slope effects, while the convolution term encodes 
global curvature ones.  The only relevant length scale is the 
saturation length $L$ of the sand flux.  The other relevant physical 
parameter is the saturated sand flux on a flat bed $Q$.  All the 
lengths are calculated in units of $L$, time in units of $Q/L^{2}$, 
and fluxes in unit of $Q$.  $A$ and $B$ could in principle be 
predicted by the Jackson and Hunt analysis but we rather take them as 
two tunable phenomenological constants.

In three dimensions, equations are very similar, albeit slightly 
different.  In order to express the total sand flux (which is now a 2D 
vector), we need to distinguish saltons and reptons \cite{H03}.  The 
reason is that in contrast to the saltons, which \emph{follow the 
wind}, the motion of the reptons is sensitive to the local slope 
\cite{H77}.  Because the reptons are dislodged by the saltons, we 
assume that their fluxes are proportional \cite{A03}, so that the 
total flux can be written as the sum of two terms, one along the wind 
direction $\vec{x}$ and the other along the steepest slope \cite{H77}: 
\be
{\vec q}_{tot} = q \, \vec{x} - D q \, \nab h.
\label{fluxtot}
\ee
The continuity equation then takes its generalized form
\be
\label{cc1bis}
\dr_t h + \nab \! \cdot {\vec q}_{tot}  =  0.
\ee
The down slope flux of reptons acts as a diffusive process.  The 
diffusion coefficient is proportional to $q$ so that no new scale is 
introduced -- $D$ is a dimensionless parameter.  This diffusion term 
introduces a non-linearity that has a slight effect only: almost the 
same dynamics is obtained if a constant diffusion coefficient is used 
instead.  Equations (\ref{cc2}) and (\ref{cc3}) can be solved 
independently in each slice along $x$.

In summary, the \ccc\ model considered here includes in a simple way
all the known dynamical mechanisms for interactions between the
dune shape, the wind and the sand transport. 

\subsection{Propagative solutions of the \ccc\ model.}
\begin{figure}[p!]
\vspace{-0.1cm} \includegraphics{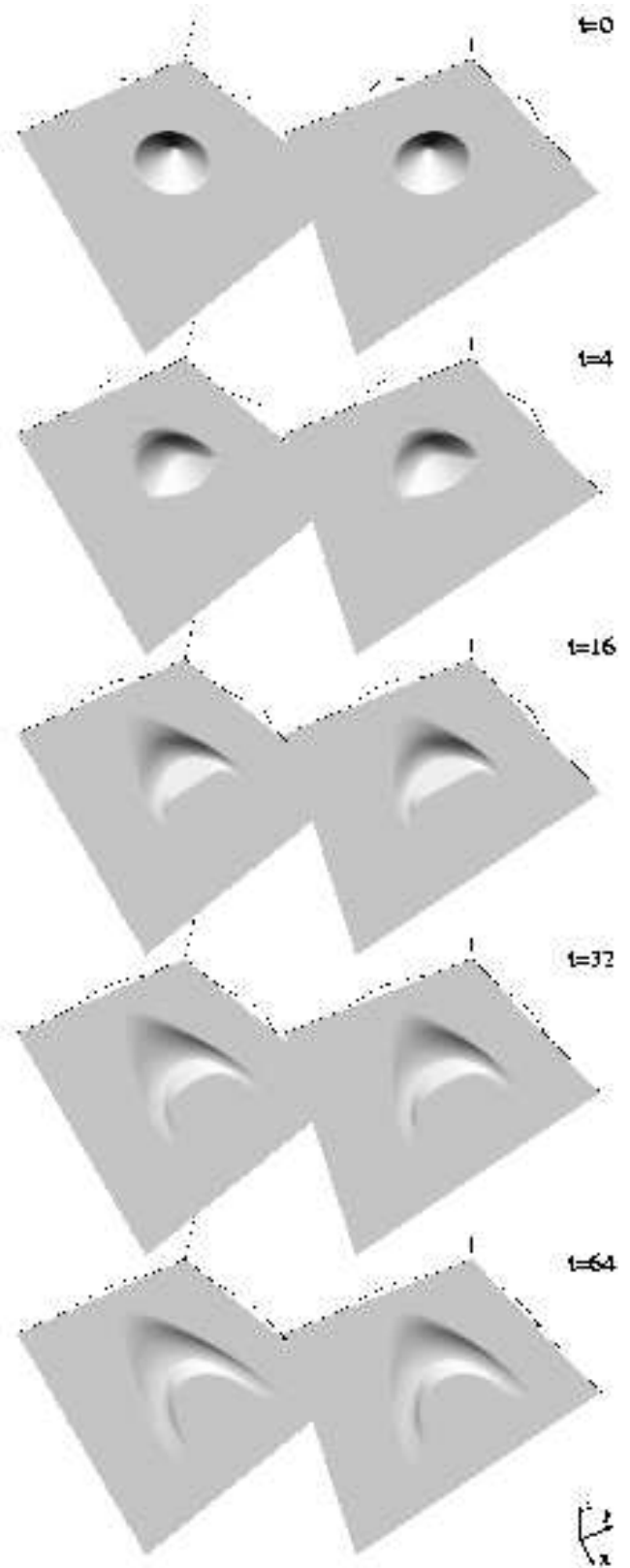} \vspace{-0.3cm}
\caption{Evolution from a conical sand pile to a steady propagative
barchan computed from the \ccc\ model.  To obtain this steady solution
the output flux is re-injected homogeneously at the upwind
boundary. Times are given in units of $L^2/Q$.  Stereoscopic view: a)
place the figure at $\sim 60~cm$ from your eyes b) focus behind the
sheet, at infinity (you should see three dunes) c) focus on the middle
dune and relax d) you should see the shape in 3D.}
\label{BarchanEv}
\end{figure}
\begin{figure}[t!]
\includegraphics{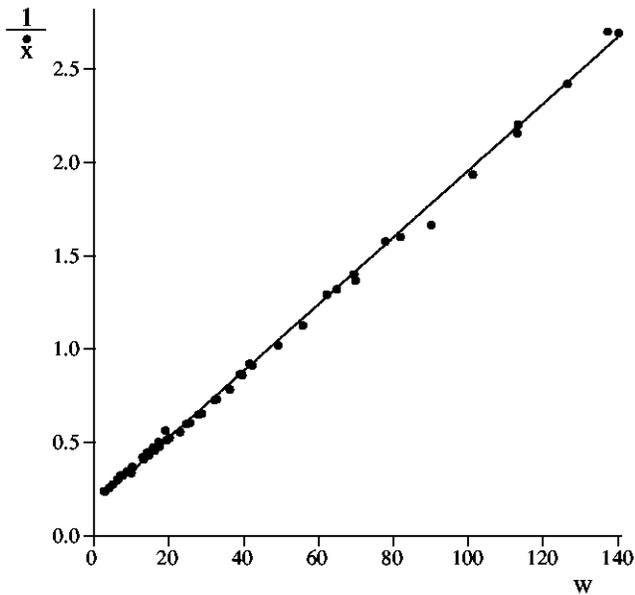} \vspace{-0.1cm}
\caption{Relationship between the inverse velocity $1/\dot x$ and the
width $w$ for barchans in the steady state.  The line corresponds to
the best fit by a Bagnold-like relation of the form $\dot x = a
Q/(w+w_c)$. It gives $a=56$ and $w_c=9.5$.}
\label{VelocityWidth}
\end{figure}

In a previous paper \cite{ACD02b}, we have studied in details steady 
propagative solutions in the 2D case.  They also apply to transverse 
dunes, {\it i.e.} invariant in the $y$ direction.  We will now focus 
on three dimensional solitary dunes computed with the \ccc\ model 
presented in the latter section.  The details of the integration 
algorithm and the numerical choice of the different parameters can be 
found in the Appendix and a more detailed discussion about the 
influence of the diffusion parameter is discussed in \cite{H03}.  
Figure~\ref{BarchanEv} shows in stereoscopic views the time evolution 
of an initial conical sand pile ($t=0$).  Horns quickly develop 
($t=16$ and $t=32$) and a steady barchan shape is reached after 
typically $t=50$.  Note that the propagation of the dune is not shown 
on figure~\ref{BarchanEv}: the center of mass of the dune is always 
kept at the center of the computation box.

The original \ccc\ model proposed by Kroy, Sauermann {\it et al.}
\cite{KSH01,S01} was the first of a long series of models in which
a steady solitary solution could be exhibited, with all the few known
properties of barchans. In particular, the dunes present a nice
crescentic shape with a length, a width, a height and a horn size that
are related to each others by linear relationships.  They propagate
downwind with a velocity inversely proportional to their size, as
observed on the field.  These properties are robust inside the class
of modeling, since we get the same results with the simplified version
that we use here.  We will only show in the following two of these
properties, important for the stability discussion, namely the
velocity and the volume as functions of the dune size.

Since they are linearly related one to the others, all the dimensions
are equivalent to parameterize the dune size.  We choose the width $w$
as it is directly involved in the expression of the sand flux at the
rear of the dune.  Figure \ref{VelocityWidth} shows the inverse of the
propagation velocity of the dune as a function of $w$.  The velocity
decreases as the inverse of the size:
\be
\dot x \sim \frac{a Q}{w+w_c} \, .
\label{propagation}
\ee
Preliminary fields measurements of the displacement of the dunes shown 
on figure \ref{CoLayoune} over $27$ years have given $aQ = 
3700~m^{2}/year$ and $w_{c}=33~m$.  The transverse velocity $\dot y$ 
is found to be null, as lateral inhomogeneities of the sand flux are 
unable to move dunes sideways \cite{LSHK02}.  The volume of $V$ is 
plotted on figure \ref{VolumeWidth}.  This relation is well fitted by: 
\be
V = b w^2 (w + w_v),
\label{volume}
\ee
where the numerical coefficients are $b\sim 0.011$ and $w_v \sim 
22.9$.  This value roughly corresponds to the volume of a half 
pyramid, with a height $h\sim 0.1 \, w$ and a width $w$ which gives a 
volume $V \sim w^{3}/60$.  One can observe that barchan dune are not 
self similar object: the deviation observed for small dunes is related 
to the change of shape due to the existence of a characteristic length 
$L$.  
\begin{figure}[t!]
\includegraphics{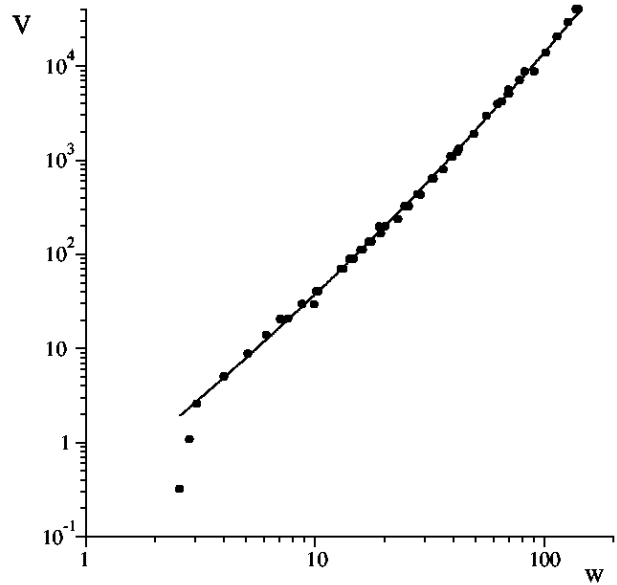} \vspace{-0.1cm}
\caption{Relationship between the volume $V$ and the width $w$ of
solitary barchans.  Note the log-log scales.  The solid line
corresponds to the best fit by the relation $V = b w^2 (w + w_v)$.
It gives $b=0.011$ and $w_v=22.9$.}
\label{VolumeWidth}
\end{figure}

\subsection{Instabilities}

The choice of the boundary conditions is absolutely crucial: to get 
stationary solutions, the sand escaping from the dune and reaching the 
downwind boundary is uniformly re-injected at the upwind one.  
Obviously, this ensures the overall mass conservation.  Doing so, the 
simulation converges to a barchan of well defined shape of width 
$w_\infty$ with a corresponding sand flux $q_\infty$.

\begin{figure*}[t!]
\includegraphics{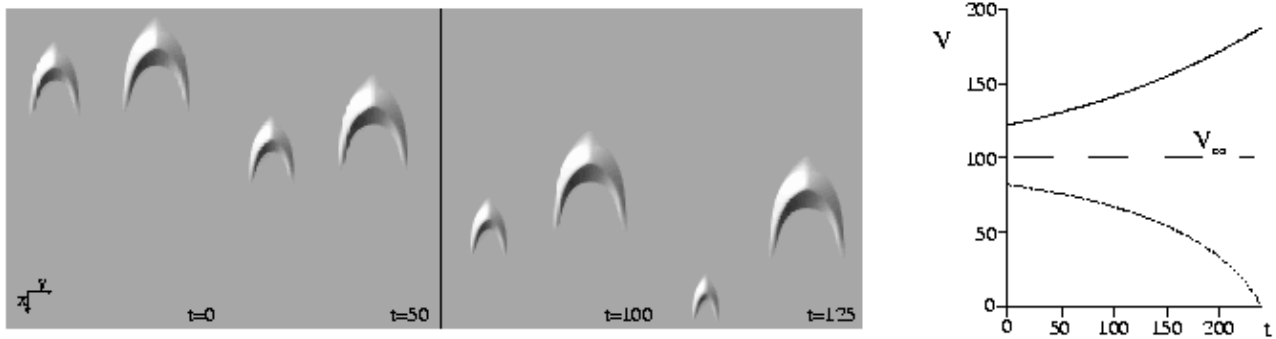} \vspace{-0.1cm} 
\caption{Origin of the flux instability.  Two dunes are submitted to a 
given flux, which is not the equilibrium sand flux for both dunes.  
The small dune (dotted line) is then under supplied an can only shrink.  On the 
contrary, the bigger dune (solid line) receive too much sand and grows.  The evolution
of their volume $V$ is depicted on the right.  Eventually, the small one 
disappears.}
\label{FluxInstabilityCCC}
\end{figure*}

However, under natural conditions, the input flux $q$ is imposed by 
the upwind dunes.  We thus also performed simulations with a given and 
constant incoming flux.  Figure~\ref{FluxInstabilityCCC} shows the 
evolution of two dunes of different sizes under an imposed constant 
input flux.  One is a bit larger than the steady dune corresponding to 
the imposed flux, and the other is slightly smaller.  It can be 
observed that none of these two initial conditions lead to a steady 
propagative dune: the small one shrinks and eventually disappears 
while the big one grows for ever.  The steady solution obtained with 
the re-injection of the output flux is therefore unstable.

If solitary dunes are unstable, it is still possible that the 
interaction between dunes could stabilize the whole field.  It is not 
what happened in the \ccc\ model. Instead, an efficient coarsening 
takes place as shown by Sauermann in the chapter $8$ of \cite{S01}.

As a first conclusion, the \ccc\ model predicts that solitary barchans 
and barchan fields are unstable in the case of a permanent wind.  We 
will see below that these two instabilities are generic and not due to 
some particularity of the modeling.  In particular, they can be also 
observed with the more complicated equations of Kroy, Sauermann {\it 
et al.} who deal with a non-linear charge equation and take explicitly 
into account the existence of a shear stress threshold to get erosion.

Therefore, we can wonder what are the dynamical mechanisms responsible 
for these instabilities. Would they have time/length to develop in an 
actual barchan field?  Seeking answers to these questions, we will now 
investigate the two instabilities in a more general framework.  As a 
first step, we will investigate the time and length scales associated 
to the evolution of barchan dunes.


\section{Time and length scales}

Three different time scales govern the dynamics of dunes: a very short
one for aerodynamic processes (i.e. the grain transport), the turnover
time for the dune motion, and a much larger time scale involved in the
evolution of the dune volume and shape under small perturbations of
the wind properties.

\subsection{Turnover time}

The dune memory time is usually defined as the time needed to
propagate over its own length.  Since the length and the width of the
dune are almost equal -- this is only a good approximation for steady
dunes -- we will use here the turnover time:
\be
\tau_t=\frac{w}{\dot x}.
\ee
In the geological community, the turnover time is believed to be the
time after which the dune looses the memory of its shape.  The idea is
that a grain remains static inside the dune during a cycle of typical
time $\tau_t$: it then reappears at the surface and is dragged by the
wind to the other side of the dune.  In other words, after $\tau_t$
all the grains composing a dune have moved, and the internal structure
of the dune has been renewed.  But this does not preclude memory of
the dune \emph{shape} at times larger than $\tau_t$, and one can
wonder whether $\tau_t$ is the internal relaxation time scale to reach
its equilibrium shape.
\begin{figure}[t!]
\includegraphics{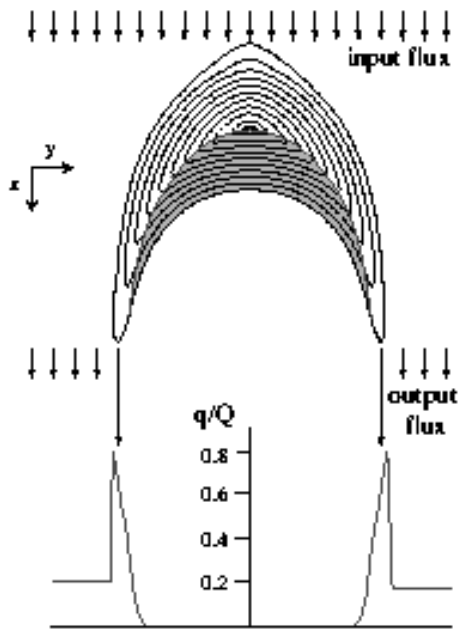} \vspace{-0.1cm} \caption{Top: three 
dimensional shape of a barchan dune obtained with the \ccc\ model, 
with semi-periodic boundary conditions to ensure the mass 
conservation.  Bottom: profile of the resulting output sand flux.  The 
sand loss is localized at the tips of the horns.  There, the flux is 
almost saturated: $q \sim Q$.}
\label{Barchan3D}
\end{figure}
The scaling (\ref{propagation}) of the propagation speed involves the 
cut-off length scale $w_c$, which can be measured by extrapolating the 
curve of \ref{VelocityWidth} to zero.  Note that the existence of a 
characteristic length scale also appears in the dune morphology 
\cite{SRPH00,KSH01,S01,ACD02b}.  In the following, we will assume that 
the barchans are sufficiently large to be considered in the asymptotic 
regime.  We checked that introducing cuts-off $w_c$ or $w_v$ to 
capture the shape of curves like that of figures \ref{VelocityWidth} 
or \ref{VolumeWidth} in the region of small $w$ does not change 
qualitatively the results.  In the following we then take $w_c=0$ and 
$w_v=0$ for simplicity.  Under this assumption, using the expression 
(\ref{propagation}) of the propagation speed, the turnover time reads:
\be
\tau_t=\frac{w^2}{aQ}.
\ee

Of course, the length scale $\lambda_t$ associated with the turnover
time is the size of the dune itself:
\be
\lambda_t=w.
\ee

\subsection{Relaxation time}
Let us consider, now, a single barchan dune submitted to a uniform 
sand flux.  The evolution of its volume is governed by the balance of 
incoming $\varphi_{in}$ and escaping $\varphi_{out}$ sand volumes per 
unit time:
\be
\dot V = \varphi_{in} - \varphi_{out}.
\ee

$\varphi_{in}$ is directly related to the local flux $q$ upwind the
dune, defined as the volume of sand that crosses a horizontal unit
length line along the transverse direction $y$ per unit time.  Assuming
that this flux $q$ is homogeneous, the dune receives an amount of sand
simply proportional to its width $w$:
\be
\varphi_{in}=q w.
\ee
\begin{figure}[t!]
\includegraphics{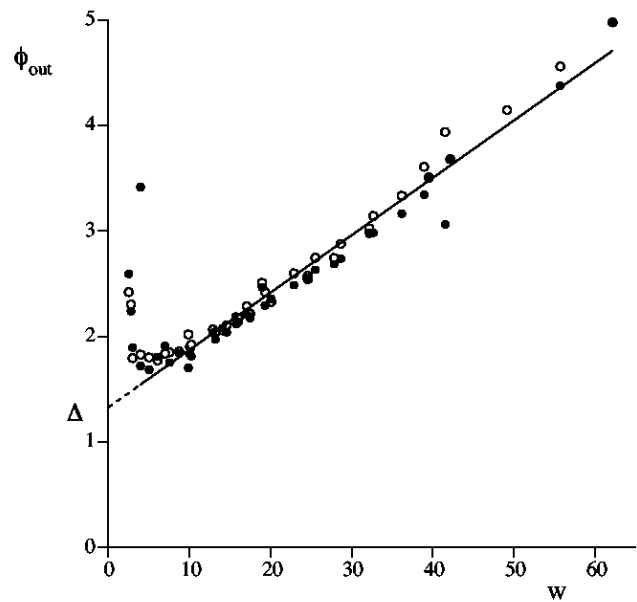} \vspace{-0.1cm} \caption{Output flux
$\varphi_{out}$ as a function of the barchan width $w$ for the
equilibrium input flux (white circles) and for a null input flux
(black circles).  $\varphi_{out}$ is not simply proportional to $w$
and does not vanish at small size.  As a consequence, the sand loss is
proportionally smaller for a large dune than for a small one.  The
straight line corresponds to the best fit by a function of the form
$\varphi_{out}=Q (\Delta+\alpha w)$.  It gives $\Delta=1.33$ and
$\alpha=0.05$.}
\label{loss}
\end{figure}

The loss of sand $\varphi_{out}$ is not simply proportional to $w$ 
because the output flux is not homogeneous.  Figure~\ref{Barchan3D} 
shows the flux in a cross-section immediately behind the dune.  One 
can see that the sand escapes only from the tip of the horns, where 
there is no more avalanche slip face.  As a matter of fact, the 
recirculation induced behind the slip face traps all the sand blowing 
over the crest.  We computed in the model the output flux 
$\varphi_{out}$ as a function of the dune width $w$ 
(figure~\ref{loss}).  Within a good approximation it grows linearly 
with $w$: 
\be
\varphi_{out} \sim Q (\Delta + \alpha w).
\ee

For the set of parameters chosen, the best fit gives $\alpha=0.05$ and
$\Delta=1.33~L$. Note that the discrepancy of $\varphi_{out}$ with the
linear variation for small dunes can be understood by the progressive
disappearance of the slip face (domes), leading to a massive loss of
sand.

It can be observed from figure~\ref{Barchan3D} that $q$ is almost
saturated in the horns.  The ratio $\varphi_{out}/Q$ then has a
geometrical interpretation as it gives an estimate of the size of the
horn tips.  Therefore, in the \ccc\ model, the horn size is not
proportional to the dune width, but grows as $\Delta + \alpha w$.
This is consistent with the observations made by Sauermann \emph{et
al.} in southern Morocco: they claim that, at least for symmetric
solitary dunes, the slip face is proportionally larger for large dunes
than for small ones, i.e. that the ratio of the horns width to the
barchan width decreases with $w$.

With these two expressions for the input and output volume rates, the 
volume balance reads: 
\be
\dot V = q w - Q (\alpha w+ \Delta).
\label{dotvolume}
\ee

If we call $w_{\infty}$ and $q_{\infty}$ the width and the flux of the 
steady dune for which the dune volume is constant ($\dot{V}~=~0$), we 
can define $\tau_{r}=(V-V_\infty)/\dot V$, taken around the fixed 
point.  We get:
\be
\tau_{r} = \frac{3 b w_\infty^3}{Q \Delta}.
\label{taur}
\ee

It also gives us the relaxation length for the dune $\lambda_{r}$, 
which is the distance covered by the dune during the time $\tau_{r}$, 
i.e:
\be
\lambda_{r} = \frac{3 ab w_\infty^2}{\Delta}.
\ee

\subsection{Flux screening length}

For a dune field, the situation is a little bit more complex.  The 
flux at the back of one dune is due to the output flux of an upwind 
dune.  The latter is strongly inhomogeneous since the sand is only 
lost by the horns (figure~\ref{Barchan3D}).  Field observations show 
that there is a sand less area downwind of the barchans -- see also the 
inset of figure~\ref{screening}.  This zone is larger than the 
recirculation bubble and indicates a small amount of sand trapped by 
the roughness of the ground.  The fact that this `shadow' heals 
up is a signature of a lateral diffusion of the sand flux.  The length 
of the shadow is typically a few times the dune width and is in 
general smaller than the distance between dunes.  So, the flux can be 
considered as homogeneous when arriving at the back of the next dune.
\begin{figure}[t]
\includegraphics{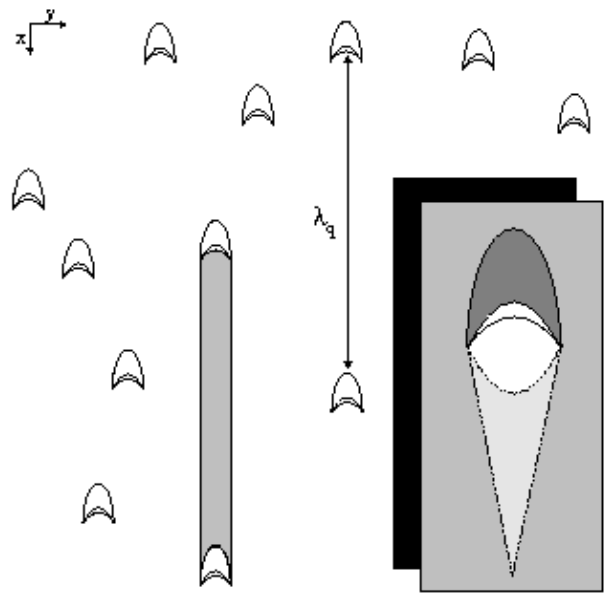} \vspace{-0.1cm} \caption{ The flux 
screening length is the mean free path along the wind direction.  In 
other words, it is the mean longitudinal distance between two dunes.  
Inset: the sand flux is much larger on the back of the dune (dark 
zone) than on the surrounding ground (gray zone).  Downwind the dune, 
it becomes inhomogeneous and in particular, it is null inside the 
recirculation bubble (white zone) and low in the triangular shadow 
zone (light grey).  After few dune sizes, the diffusion of grains 
rehomogenizes the flux.}
\label{screening}
\end{figure}

The distance $\lambda_{q}$ over which the flux changes is thus the 
distance, along the wind direction, between two dunes.  It is the mean 
free path of one grain traveling in straight line along the wind 
direction.  Let us consider an homogeneous dune field composed of 
identical dunes of width $w_{\infty}$.  The number of dunes per unit 
surface is $N_{\infty}$.  It can be inferred from figure 
\ref{screening} that on average there is one dune in the surface 
$\lambda_{q} w_{\infty}$ (colored in gray on figure~\ref{screening}).  
The flux screening length thus depends on the density of dunes as: 
\be
\lambda_{q} = \frac{1}{N_{\infty} w_{\infty}}.
\label{screenlength}
\ee

Note that this length is larger than the average distance 
$(N_{\infty})^{-1/2}$ between dunes -- just like the mean free path in 
a gas.  Since the grains in saltation on the solid ground go much 
faster than the dune (by more than five orders of magnitude), the flux 
screening time $\tau_{q}$ can be taken as null: 
\be
\tau_{q}=0.
\ee

\subsection{Orders of magnitude}
These different time scales can be estimated using the orders of 
magnitude obtained from field observations in the region of 
figure~\ref{CoLayoune}.  The velocity/width relationship has allowed 
to estimate $aQ \sim 3700~m^{2}/year$ and $w_{c} \sim 33~m$ in this 
region.  Combined with \ccc\ model results, this gives estimates of the 
saturation length $ L =~3.5 m$, the minimal horn width $\Delta =4.6~m$ 
and the saturated flux far from any dune $Q \sim 66~m^{2}/year$.  
These values are corroborated by direct measurements of these 
quantities \cite{B41,ACD02a}.

Let us consider a small dune of width $20~m$ and a large dune of width 
$100~m$ belonging to the corridors of dunes shown on 
figure~\ref{CoLayoune}.  The distance $\lambda_r$ covered by the dune 
before the equilibrium between the size and the sand flux be reached 
is respectively $160~m$ and $4~km$.  In all the cases, it is much 
smaller than the dune field extension (typically $100~km$ 
corresponding to $5000$ small dune widths or $1000$ large dune 
widths).  Obviously, $\lambda_r$ is much larger than the turnover 
length $\lambda_t=w$, and it is therefore clear that the turnover 
scales does not represent the memory of the dune.

The density of dunes can be inferred from figure~\ref{CoLayoune} and 
is around $0.1~{/w^2_\infty}$ (the average distance between dunes is 
around $3$ dune sizes).  Directly from figure~\ref{CoLayoune} or from 
formula (\ref{screenlength}), the flux adaptation length $\lambda_q$ 
is around $10$ dune sizes i.e.  $200~m$ for the small dune and $1~km$ 
for the large one.  Obviously, $\lambda_{q}$ can be very different 
from place to place.  For instance, the left corridor shown on 
figure~\ref{CoLayoune} is much denser than the third from the left.  
If the density of small dunes is $1~{/w^2_\infty}$ instead of 
$0.1~{/w^2_\infty}$, $\lambda_q$ becomes equal to the dune size 
($20~m$).

Using the previous value of $Q$, the dune velocities are $180~m/year$ 
and $37~m/year$ for the $20~m$ and $100~m$ barchans respectively.  The 
corresponding turnover times $\tau_t$ are $5.2~weeks$ and $2.7~years$, 
while the relaxation time $\tau_r$ is as large as $10~months$ for 
the small dunes and $1.1~century$ for the large ones.  Finally, the 
flux adaptation time $\tau_{q}$ is equal to the flux screening length 
$\lambda_{q}$ divided by the grain speed ($\sim 1~m/s$).  It can thus 
be estimated to $3~minutes$ for $20~m$ barchans and $16~minutes$ for 
the $100~m$ ones.

The scale separation of the three times is impressive.  $\tau_{q} \sim 
3~minutes \ll \tau_{t} \sim 5.2~weeks \ll \tau_{r} \sim 10.4~months$ 
for the small dunes, whereas for the large ones it reads $\tau_{q} 
\sim 16~minutes \ll \tau_{t} \sim 2.7~years \ll \tau_{r} \sim 
1.1~century$.  This shows that the annual meteorological fluctuations 
(wind, humidity) have potentially important effects: the actual memory 
time is always larger than seasonal time.  

Sauermann has estimated a characteristic time for the evolution of the 
volume of a $100~m$ wide dune.  He found several decades \cite{S01}, 
which is comparable to the value we found for $\tau_{r}$.  On this 
basis he concluded that `` considering this timescale it is justified 
to claim that barchans in a dune field are only marginally unstable.''  
This is a misleading conclusion as the length $\lambda_{r}$ ($4~km$ 
for $w = 100~m$) should be compared to the corridor size ($\sim 
100~km$).  Moreover, for small dunes as those on the left corridor of 
figure \ref{CoLayoune}, $\lambda_{r}$ is found to be as small as 
$160~m$ which is of the order of one hundredth of the portion of field 
displayed on the photograph.  The evolution time and length scales 
could be thought as `very large' Ñ- with respect to human scales -- 
but compared to the dune field size, they turn out to be small.  
Therefore, barchans have in fact the time and space to change their 
shape and volume along the corridors.


\section{Flux instability}
\subsection{Stability of a solitary barchan}
We seek to understand the generality of the instabilities revealed by
the \ccc\ model.  We will first investigate theoretically the
stability of a solitary dune in a constant sand flux $q$.  We recall
that the overall volume of sand received by this dune per unit time is
simply proportional to its width: $\oldphi_{in} = q w$.  As found in
the \ccc\ model, we suppose that $\oldphi_{out}=Q (\Delta + \alpha w)$
(figure~\ref{FluxInstability} left).  Let us investigate what happens
for different values of the input flux $q$.

If $q < \alpha$ (dot-dashed line) the two curves $\oldphi_{in}(w)$ and 
$\oldphi_{out}(w)$ do not cross, which means that no steady solution 
can be found.  Since the input sand volume rate is too low, any dune 
will shrink and eventually disappear.  On the other hand, a fixed 
point $w_{\infty}$ does exist for $q > \alpha$ (thin solid line).  
Suppose that this dune is now submitted to a slightly larger (resp.  
smaller) flux $q$ (dotted lines): it will grow (resp.  shrink).  
However, the corresponding steady states are respectively smaller and 
larger, so that they cannot be reached dynamically.  We now fix the 
input flux to $q_{\infty}$ and change the dune size $w$, as in 
figure~\ref{FluxInstabilityCCC}.  A dune of width slightly smaller 
than $w_\infty$ under will shrink more and more because it looses sand 
more than it earns.  In a similar way, a dune larger than $w_\infty$ 
will ever grow.  In other words, the steady solutions are unstable.

This mechanism explains the flux instability of \ccc\ barchans.  This
stability analysis is in fact robust and not specific to the linear
choice for $\oldphi_{out}$.  Any more complicated function would lead
to the same conclusion provided that $\oldphi_{in}$ crosses
$\oldphi_{out}$ from below.  The stability only depends on the
behavior of the $\oldphi$'s in the neighborhood of the steady state.

How could a solitary barchan be stable?  It is enough that 
$\oldphi_{in}$ crosses $\oldphi_{out}$ from above.  Without loss of 
generality, we can keep a linear dependence of $\oldphi_{out}$ on $w$ 
in the vicinity of the fixed point, but this time with $\Delta <0$ 
(figure~\ref{FluxInstability} right).  In this case, the situation for 
which the input sand flux $q$ is larger than $\alpha$ (dot-dashed line) 
leads to an ever growing dune.  Steady solutions exist when $q < 
\alpha$.  Because a smaller (resp.  larger) sand flux now corresponds 
to a smaller (resp.  larger) dune width, these solutions are, by 
contrast, stable.  
\begin{figure}[t!]
\includegraphics{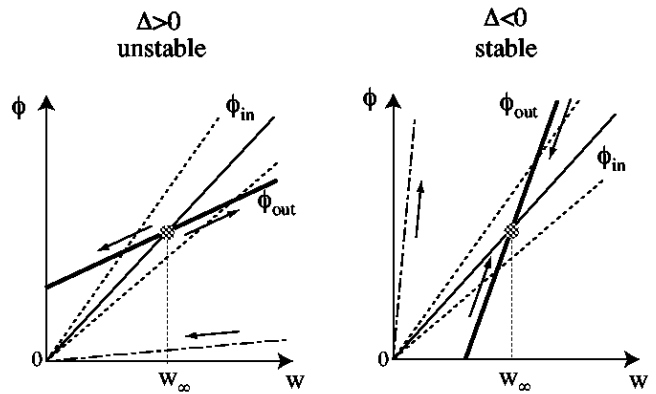} \vspace{-0.1cm}
\caption{The output volume rate $\oldphi_{out}=Q(\Delta + \alpha w)$
gives,
when compared to the input rate $\oldphi_{in} = q w$, steady solutions
that are unstable if $\Delta>0$, and stable if $\Delta<0$.  The main
point is to see whether the two lines cross from below or above at the
steady point.}
\label{FluxInstability}
\end{figure}

In a more quantitative and formal way, the mass balance for a barchan
(\ref{dotvolume}) can be rewritten in terms of the dune width only:
\be
\dot w=\frac{q_\infty w - Q (\Delta+\alpha  w)}{3b w^2}.
\label{dotw}
\ee
Linearizing this equation around the fixed point $w_\infty$ we obtain:
\be
\tau_r \dot w = w-w_\infty.
\ee
The sign of the relaxation time $\tau_r$ is that of $\Delta$ -- see 
relation (\ref{taur}).  Therefore if $\Delta$ is positive, $w$ will 
quickly depart from its steady value $w_\infty$.  In the inverse case 
$\Delta<0$, any deviation of $w$ will be brought back to $w_\infty$.  
\begin{figure}[t]
\includegraphics{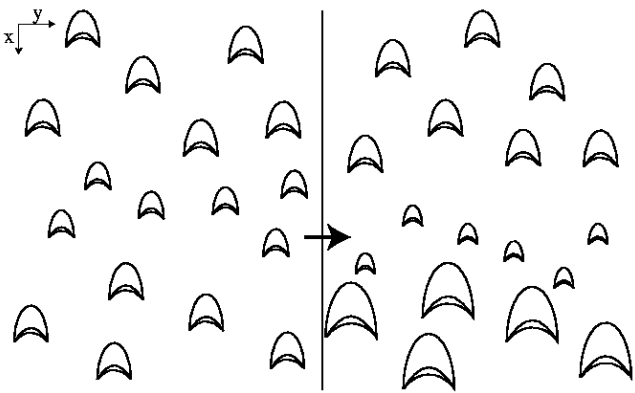} \vspace{-0.1cm}
\caption{Sketch showing the instability due to the exchange of mass
between the dunes: dunes slightly smaller than $w_{\infty}$ loose
sand and make their downwind neighbors grow. Note that the dune field
is assumed to remain locally homogenous.}
\label{viaflux}
\end{figure}

In summary, the stability of a solitary barchan depends whether the 
ratio of the output volume rate to its width $\oldphi_{out}/w$ 
increases or decreases with $w$.  This quantity is perhaps not easy to 
measure on the field but we have shown that it is directly related to 
the ratio of the size of the horn tips to the dune width.  If viewed 
from the face, the horn tips become in proportion smaller as the dune 
size increases, the barchan is unstable.  This is what is predicted by 
the \ccc\ model, in agreement with the few field observations 
\cite{SRPH00}.

\subsection{Stability of a dune field}

At this point, the stability analysis leads to the fact that a single
solitary barchan is unstable.  Could then dunes be stabilized by their
interaction via the sand flux?  Let us consider a dune field which is
locally homogeneous and composed, around the position ($x$,$y$),
barchans of width $w$ with a density $N$.  We ignore for the moment
the fact that dunes can collide.  The conservation of the number of
dunes then reads
\be
\dot N=\dr_t N + \dr_x (\dot x N) + \dr_y (\dot y N)= 0.
\ee
The eulerian evolution of the dunes width is given by:
\be
\dot w=\dr_t w + \dot x \dr_x w+ \dot y \dr_y w.
\ee
This equation has to be complemented by the equation governing the
evolution of the sand flux $q$ between the dunes which results
from the variation of the volume of these dunes (and reciprocally):
\be
\dr_x q = - N \dot V.
\ee
These three equations can be closed using the previous modeling of
$\dot x$, $\dot y = 0$, $\dot w$ and $\dot V = 3bw^{2}\dot{w}$.  Any
homogeneous field of barchans of width $w_\infty$ and density
$N_{\infty}$ is a solution, provided that there is a free flux
$q_\infty = Q (\Delta/ w_\infty+\alpha)$ between the dunes.
\begin{figure}[t!]
\includegraphics{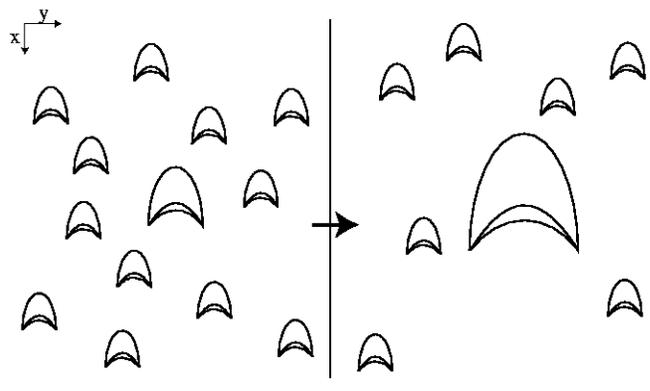} \vspace{-0.1cm} \caption{Sketch showing
the instability due to the collisions between the dunes.  If one dune
is slightly larger than the others, it goes slower and will absorb
incoming dunes.}
\label{viacol}
\end{figure}

We are now interested in the stability of this solution towards
locally homogeneous disturbances.  Note that such a choice is still
consistent with equations that do not take collisions into account.
We thus expand $w, q$ and $N$ around their stationary values
$w_{\infty}, q_{\infty}, N_{\infty}$, and introduce the length and
time scales $\lambda_r, \lambda_q$ and $\tau_r$.  We get:
\bea
\tau_r \dr_t N & + & \lambda_r \dr_x N -
\frac{N_{\infty}}{w_{\infty}} \lambda_r \dr_x w = 0, \label{dvptN} \\
\tau_r \dr_t w & + & \lambda_r \dr_x w = (w - w_{\infty}) +
\frac{w_\infty^2}{Q \Delta} (q - q_{\infty}), \label{dvptwbar}\\
&&\lambda_q \dr_x q = - (q - q_{\infty})
-\frac{Q\Delta}{w_\infty^2} ( w - w_{\infty}).
\label{dvptq}
\eea
Without loss of generality we can write the disturbances under the 
forms: $q - q_{\infty} = q_{1} e^{\sigma t + i k x}$, $ w - w_{\infty} 
= w_{1} e^{\sigma t + i k x}$ and $N - N_{\infty} = N_{1} e^{\sigma t 
+ i k x}$.  Solving the system of linear equations we obtain the 
expression of the growth rate $\sigma$ as a function of the wavenumber 
$k$: 
\be
\tau_r \sigma = ik \left ( \frac{\lambda_q}{1+(k\lambda_q)^2} -
\lambda_r \right ) + \frac{(k\lambda_q)^2}{1+(k\lambda_q)^2}.
\ee
The sign of the real part of the growth rate $\sigma$ is that of
$\tau_r$ and thus of $\Delta$.  The stability of the dune field is
therefore that of the solitary dune. If $\Delta<0$, which means that
all the individual dunes are stable, the field is (for obvious reasons)
stable. But in fact, all the individual dunes are unstable
($\Delta>0$), so that a field in which dunes interact via the
sand flux is also unstable.

The result of the above formal demonstration, can also be understood 
via a simple argument illustrated on figure~\ref{viaflux}.  Consider a 
barchan dune field at equilibrium: for each dune, input and output 
volume rate are equal.  Now, imagine that the input flux of a dune 
slightly decreases for some reasons.  As explained in the previous 
subsection, if this dune is unstable ($\Delta>0$) it tends to shrink.  
Consequently its output flux increases, and makes its downwind 
neighbors grow.  Therefore, even a small perturbation of the sand flux 
can dramatically change the structure of the field downwind.

\section{Collisional instability}

The free flux is not the only way barchans can influence one another.
If sufficiently close, they can interact through the wind, i.e.
aerodynamically.  This is possible when the dunes get close to each
other.  In this case, they actually collide.  We therefore would like
to investigate the behavior of one particular dune in the middle of
the field.

Let us consider a homogeneous field of barchans of width $w_{\infty}$, 
with an additional dune of size $w=(1+\eta) w_\infty$.  The variation 
of the volume of this dune is due to the sand flux as well as the 
collisions of incoming dunes.  These collisions are a direct 
consequence of the fact that smaller dunes travel faster 
(Eq.~\ref{propagation}).  The number of collisions per unit time is 
proportional to the dune density $N_\infty$ times the collisional 
cross section $w+w_\infty$ times the relative velocity 
$aQ~(1/w_\infty-1/w)$.  We assume that the collisions lead to a 
merging of the two dunes.  Then, each collision leads to an increase 
of the mass of the larger dune by $V_\infty=b w_\infty^3$.  We can 
then write for this particular dune: 
\bea
\dot V = q w & - & Q (\alpha w + \Delta) \nonumber \\
& + & N_\infty V_\infty (w+w_\infty) \frac{aQ (w-w_\infty)}{w w_\infty}.
\eea
Introducing a critical dune density $N_c$ as
\be
N_c=\frac{-\Delta}{2ab w_\infty^3},
\ee
the equation governing the evolution of the width of the dune
considered reads:
\be \tau_r \dot{\eta} = \left (1 - \frac{(2+\eta)}{2(1+\eta)}
\frac{N_\infty}{N_c} \right ) \frac{\eta}{(1+\eta)^2}.
\ee
Note that, rigorously speaking, $N_c$ is a positive quantity and thus 
a true density only for $\Delta<0$ (see below).  Figure~\ref{sigma} 
shows $\dot{\eta}$ as a function of the dune size.  Expanding linearly 
around $\eta=0$, we obtain the growth rate $\sigma=\dot{\eta}/\eta$ 
as: 
\be
\tau_r \sigma= 1-\frac{N_\infty}{N_c}.
\ee
\begin{figure}[t]
\includegraphics{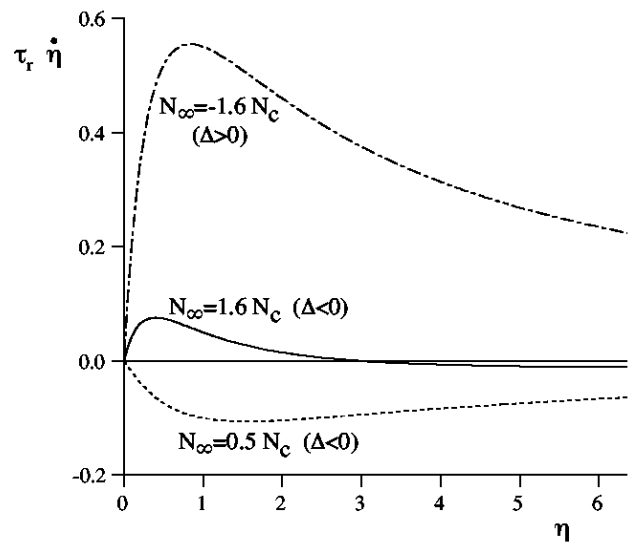} \vspace{-0.1cm} \caption{Growth rate
$\tau_{r} \dot{\eta}$ of a dune due to collisions, as a function of
the rescaled size $\eta$ in three different cases.  If the solitary
dune is unstable ($\Delta>0$), the field is also unstable towards the
collisional instability (dot-dashed line).  If the solitary dune is
stable ($\Delta<0$), the stability of the field depends on the dune
density.  At high density (solid line), the field is linearly unstable
while at low density (dashed line), it is stable towards any
disturbance.}
\label{sigma}
\end{figure}

Therefore, if the dunes are individually unstable (case $\Delta>0$, 
figure~\ref{sigma} a), the dunes are always unstable towards the 
collisional instability.  The barchan field quickly merge into one big 
barchan dune.  If the dune are individually stable ($\Delta<0$), the 
same instability develops but only when the dune density is larger 
than the critical dune density $N_c$ (figure~\ref{sigma} b).  Suppose 
indeed that one collision occurs in the middle of an homogeneous 
field, creating a dune of twice its original volume.  Since it is 
larger, this dune slows down and a second collision occurs before the 
large dune has recovered its equilibrium.  If now the dune density is 
small (figure~\ref{sigma} c), the time before a second collision 
happens is sufficiently large to allow the large dune to recover its 
equilibrium volume, and in this case a field of stable dunes is stable 
towards the collision process.

\section{Barchans corridors, an open problem}
The aim of this conclusion is twofold.  We will first give an overview 
of the different results presented in this paper.  Then we will 
discuss the problem of the size selection and the formation of 
barchans corridors.

The starting point of the present work is the observation that barchan 
dunes are organized in fields stretched along the dominant wind 
direction (figure~\ref{CoLayoune}).  These barchans corridors are 
quite homogeneous in size and in spacing.  For instance, the barchans 
field between Tarfaya and Laayoune presents the same five coherent 
corridors over at least $100~km$.  This size selection is of course 
not to be taken in a strict sense: there are large fluctuations from 
one dune to another, which have {\it also} to be explained.

We have shown that the stability of a solitary dune essentially 
depends on the relationship between the size of the horns and that of 
the dune.  Indeed, the dune receives at its back a sand flux 
proportional to its size but releases sand only by its horns.  If the 
size of the horns is proportionally smaller for large dunes than for 
small ones, the steady state of the dune is unstable: it either grows 
or decay (figure~\ref{viaflux}).  If, on the contrary, the sand leak 
increases faster than the dune size, it pulls the dune back to 
equilibrium.  Furthermore we have shown that the fact that a dune is 
fed by the output flux of the dunes upwind does not change the 
stability analysis.  This is essentially because a dune can influence 
another dune downwind through the flux but there is no feedback 
mechanism.

We are thus left with a secondary question: how does the horns width
evolve with the dune size?  The only field measurements from which
the horns size can be extracted are the shape measurements of eight
dunes by Sauermann \emph{et al.} \cite{SRPH00}. The sum of the width of
the two horns is found to be between $12$ and $28~m$ for the five
small dunes they measured ($2$ and $3~m$ high), and between $12$
and $17~m$ for the three larger ones (heights between $6$ and
$8~m$).  If we trust the relevance of their selection of dunes,
this means that the horn size is almost independent of that of the
dune. This is also coherent with their claim that the slip face is
proportionally smaller and the horns larger for small dunes than
for large ones. In that case, solitary  barchans should be individually
unstable.

The second indication is provided by the \ccc\ modeling, with which we 
recover that this steady state is in fact unstable 
(figures~\ref{FluxInstabilityCCC}~and~\ref{loss}), in fact for the 
very same reasons as above.  The solution can be artificially 
stabilized by putting at the back of the dune exactly what it looses 
by its horns but this is only a numerical trick.  What determines the 
size of the horns in the model?  The 3D solutions can be thought of 
coupled 2D solutions \cite{ACD02b}.  Then, the horns start when there 
is no slip face, i.e.  when the length becomes of the order of 
the minimal size of dunes.  This simple argument leads to think that 
the horns should keep a characteristic size of order of few saturation 
lengths $L$ whatever the dune size is.  
\begin{figure}[t!]
\includegraphics{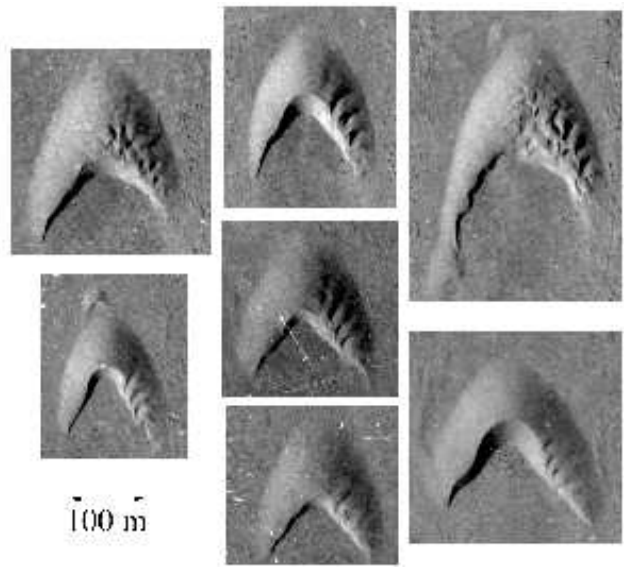} \caption{Aerial photographs of several 
barchan dunes in the same region as the field of 
figure~\ref{CoLayoune}.  They all exhibit an instability on the left 
side, leading to a periodic array of small slip-faces.}
\label{InstaBras}
\end{figure}

We have shown that there is a second robust mechanism of instability.  
We know from field measurements, numerical models and theoretical 
analysis that the dune velocity is a decreasing function of its size.  
The reason is simply that the flux at the crest is almost independent 
of the dune size and will make a small dune propagate faster than a 
large one.  This is sufficient to predict the coarsening of a dune 
field: because they go faster, small dunes tend to collide the large 
ones making them larger and slower\ldots This collision instability 
should also lead to an ever growing big dune.

The scales over which all instabilities develop are the relaxation 
time $\tau_r$ and length $lambda_r$.  For the eastern corridor of 
figure~\ref{CoLayoune}, the order of magnitude of the dune width is 
$100~m$ which gives $\tau_r \sim 1.1~century$ and $\lambda_r \sim 
4~km$.  For the western corridors, the dunes are smaller ($w\sim 20~m$ 
and the characteristic scales become $\tau_r \sim 10~months$ and 
$\lambda_r \sim 160~m$.  These lengthes are much smaller than the 
extension of the corridor ($300~km$) so that these instabilities have 
sufficient space to develop.

However, the actual barchan corridors are homogenous and one cannot 
see any evidence of such instabilities.  As a conclusion, the dune 
size selection and the formation of barchan corridors are still open 
problems in the present state of the art.  There should exist another 
robust dynamical mechanism leading to an extra-leak of the dunes, to 
balance the collisional and the flux instabilities.  There are already 
two serious candidates for this mechanism.  First, we do not have any 
information on the collision process.  In the \ccc\ model presented 
here, the collision of two dunes leads to a merging into a larger 
dune, but in the reality, it could lead to the formation of several 
dunes.  Second, we have only investigated here the case of a permanent 
wind.  We have shown that the dunes characteristic times are larger 
than one year so that the annual variations of the wind regime could 
have drastic effects.  Figure~\ref{InstaBras} shows aerial photographs 
of eight barchan dunes.  They all present an instability on their left 
side, leading to the formation of a periodic array of $1~m$ high slip 
faces.  This asymmetry suggests that it is due to a secondary wind 
(probably a storm) coming from the west (from the left on the figure).  
This instability could lead to a larger time averaged output flux than 
expected for a permanent wind.  Further work in that direction will 
perhaps shed light on the formation of nearly homogeneous corridors of 
barchan dunes.


\centerline{\rule[0.1cm]{3cm}{1pt}}
The authors wish to thanks S.~Bohn, L.~Quartier, B.~Kabbachi and Y.~Couder for
many stimulating discussions.

\vspace*{2.5cm}

{\bf \centerline{Appendix: the 3D \ccc\ model}}

\vspace*{0.5cm}

The three starting equations of the model are the conservation of
matter, the charge equation and the coupling between the saturated
flux and the dune shape $h$:
\bea
\dr_t h + \dr_x q &=& D \, \nab \cdot (q \nab h),
\label{ccc1} \\
\dr_x q &=& \frac{q_{sat} -q}{L},
\label{ccc2} \\
\frac{q_{sat}}{Q} &=& 1 +
A \int \!\! \frac{d\chi}{\pi (\chi-x)} \, \dr_x h_e +
B \, \dr_x h_e.
\label{ccc3}
\eea
We recall that the overall flux is the sum of $q$ along the wind
direction, plus an extra flux due to reptons along the steepest
slope.  The two last equations do not contain any $y$ dependence and
can thus be solved for each slice in $x$ independently, using a
discrete scheme in space ($dx$).  The conservation of matter
(\ref{ccc1}) couples the slices through the diffusion term and is solved
by a semi-implicit scheme of time step $dt$.  To speed up the numerical
computation of the saturated flux, we use the discrete Fourier transform
${\cal F}$ of the dune envelope $h_e$:
\be q_{sat} = Q\left( 1 + {\cal F}^{-1}\left\{ {\cal F}(h_e) (A|k| +
iBk) \right\}\right).
\ee
This envelope is composed of the dune profile $h(x)$ up to the point
where the turbulent boundary layer separates:
\be
x<x_b \,:   h_{e}(x) = h(x).
\ee
In the absence of any systematic and precise studies on this
separation bubble, we assume that the separation occurs when the slope
is locally steeper than a critical value $\mu_b = 0.25$:
\be
h(x_b)-h(x_b+dx)>\mu_b dx.
\ee
When the dune presents a slip face, the boundary layer thus separates
at the crest. The separation streamline is modeled as a third order
polynomial:
\be
x_b < x < x_r \,:  h_{e}(x) = a + bx +cx^{2}+dx^{3}.
\ee
The four coefficients are determined by smooth matching conditions:
\bea
h_{e}(x_b) = h(x_b), &\quad& h_{e}(x_b-dx)  = h(x_b-dx),  \\
h_{e}(x_r) = h(x_r), &\quad& h_{e}(x_r+dx)  = h(x_r+dx),
\eea
and the reattachment point $x_r$ is the first mesh point for which the
slope is nowhere steeper than $\mu_b$.  There is no grain motion
inside the recirculation bubble, so that the charge equation should be
modified to $\dr_x q = - q$ for $x_b < x < x_r$.  Similarly, on the
solid
ground ($h=0$) no erosion takes place, so that $\dr_x q = 0$.

The last important mechanism is the relaxation of slopes steeper than 
$\mu_{d}$ by avalanching.  Rather than a complete and precise 
description of avalanches of grains, we treat them as an extra flux 
along the steepest slope:
\be
\dr_t h + \dr_x q = \nab \cdot [(Dq+E \delta \mu) \nab h],
\ee
where $\delta \mu$ is nul when the slope is lower than $\mu_{d}$ and 
equal to $\delta \mu=|\nab h|^2-\mu_d^2$ otherwise.  For a 
sufficiently large coefficient $E$, the result of this trick is to 
relax the slope to $\mu_d$, independently of $E$.  Note that, as the 
diffusion of reptons, these avalanches couple the different 2D slices.  
The value of the parameters have been chosen to reproduce the 
morphological aspect ratios and are given in the following table:
\begin{center}
\begin{tabular}{ll}
\hline \hline
$A = 4.7$               & curvature effect \\
$B = 5.0$               & slope effect \\
$D  = 0.1$              & Lateral diffusion \\
$\mu_b = 0.25$          & separation slope \\
$E = 50$                & avalanches \\
$\mu_d = 0.5$           & avalanche slope \\
$dx = 0.25  - 1.0$      & grid step\\
$dt= 0.001 - 0.1$       & time step\\
$M  =  64 - 512$        & box size\\
\hline \hline
\end{tabular}
\end{center}
\noindent
The results presented in this paper have been obtained for different
discretization time and space steps, different box sizes and different
total times. This explains the slight dispersion of the measurements.


\begin{thebibliography}{}

\bibitem{B41}
R.A. Bagnold, {\it The physics of blown sand and desert dunes},
Chapman and Hall, London (1941).

\bibitem{O64}
P.R. Owen, Saltation of uniform grains in air, J. Fluid. Mech.  {\bf
20}, 225-242 (1964).

\bibitem{H77}
A.D. Howard, Effect of slope on the threshold of motion and its
application to orientation of wind ripples, Bulletin Geological
Society of America {\bf 88}, 853-856 (1977).

\bibitem{S85}
M. S\o rensen, Estimation of some aeolian saltation transport
parameters from transport rate profiles, in {\it International
workshop on the physics of blown sand}, Barndorff-Nielsen, Moller,
Rasmussen and Willets eds.  University of Aarhus, 141-190 (1985).

\bibitem{JS86}
J.L. Jensen and M. Sorensen, Estimation of some aeolian saltation
transport parameters: a re-analysis of Williams data, Sedimentology
{\bf 33}, 547-558 (1986).

\bibitem{AH88}
R.S. Anderson and P.K. Haff, Simulation of aeolian saltation, Science
{\bf 241}, 820-823 (1988).

\bibitem{AH91}
R.S. Anderson and P.K. Haff, Wind modification and bed response during
saltation of sand in air, Acta Mechanica [Suppl]{\bf 1}, 21-51 (1991).

\bibitem{ASW91}
R.S. Anderson, M. S\o rensen and B.B. Willetts, A review of recent
progress in our understanding of aeolian sediment transport, Acta
Mechanica [Suppl]{\bf 1}, 1-19 (1991).

\bibitem{S91}
M. S\o rensen, An analytic model of wind-blown sand transport, Acta
Mechanica [Suppl]{\bf 1}, 67-81 (1991).

\bibitem{WMR91}
B.B. Willetts, J.K. McEwan and M.A. Rice, Initiation of motion of
quartz sand grains, Acta Mechanica [Suppl] {\bf 1}, 123-134 (1991).

\bibitem{RM91}
K.R. Rasmussen, H.E. Mikkelsen, Wind tunnel observations of aeolian
transport rates, Acta Mechanica [Suppl]{\bf 1}, 135-144 (1991).

\bibitem{NHB93}
P. Nalpanis, J.C.R. Hunt and C.F. Barrett, Saltating particles over
flat beds, J. Fluid Mech. {\bf 251}, 661-685 (1993).

\bibitem{IR94}
J.D. Iversen and K.R. Rasmussen, The effect of surface slope on
saltation threshold, Sedimentology {\bf 41}, 721-728 (1994).

\bibitem{RIR96}
K.R. Rasmussen, J.D. Iversen and P. Rautaheimo, Saltation and wind
flow interaction in a variable slope wind tunnel, Geomorphology {\bf
17}, 19-28  (1996).

\bibitem{RVB00}
F. Rioual, A. Valance and D. Bideau, Experimental study of the
collision process of a grain on a two-dimensional granular bed, Phys.
Rev.  E {\bf 62}, 2450-2459 (2000).

\bibitem{SKH01}
G. Sauermann, K. Kroy and H.J. Herrmann, A continuum saltation model
for sand dunes, Phys.  Rev.  E {\bf 64}, 031305 (2001).

\bibitem{A03}
B. Andreotti, A two species model of aeolian sand tranport,
submitted to J. Fluid Mech. (2003).

\bibitem{B10}
H.J.L. Beadnell, The sand-dunes of the Libyan desert, Geographical
Journal {\bf 35}, 379-395 (1910).

\bibitem{F59}
H.J. Finkel, The barchans of southern Peru, Journal of Geology
{\bf 67}, 614-647 (1959).

\bibitem{C64}
A. Coursin, Bulletin de l'IFAN {\bf 3}, 989-1022 (1964).

\bibitem{LS64}
J.T. Long and R.P. Sharp, Barchan dune movement in imperial valley,
California, Geological Society of America, {\bf 75}, 149-156 (1964).

\bibitem{H67}
S.L. Hastenrath, The barchans of the Arequipa Region, Southern Peru,
Zeitschrift f\"ur Geomorphologie {\bf 11}, 300-331 (1967).

\bibitem{N66}
R.M. Norris, Barchan dune of imperial valley, California, J. Geol.
{\bf 74}, 292-306, (1966).

\bibitem{H87}
S.L. Hastenrath, The barchans of Southern Peru revisited, Zeitschrift
f\"ur Geomorphologie {\bf 31-2}, 167-178 (1987).

\bibitem{S90}
M.C. Slattery, Barchan migration on the Kuiseb river delta, Namibia,
South African Geographical Journal {\bf 72}, 5-10 (1990).

\bibitem{HH98}
P.A. Hesp and K. Hastings, Width, height and slope relationships and
aerodynamic maintenance of barchans, Geomorphology {\bf 22}, 193-204,
(1998).

\bibitem{SRPH00}
G. Sauermann, P. Rognon, A. Poliakov and H.J. Herrmann, The shape of
barchan dunes of southern Morocco, Geomorphology {\bf 36}, 47-62
(2000).

\bibitem{ACD02b}
B. Andreotti, P. Claudin and S. Douady, Selection of dune shapes and
velocities.  Part 2: A two-dimensional modelling, Eur.  Phys.  J. B
\textbf{28}, 341-352 (2002).

\bibitem{HDA02}
P. Hersen, S. Douady and B. Andreotti, Relevant lengthscale of barchan
dunes, Phys. Rev. Lett. {\bf 89}, 264301, (2002).

\bibitem{M88}
K.R. Mulligan, Velocity profiles measured on the windward slope of a
transverse dune, Earth surface processes and Landforms {\bf 13},
573-582 (1988).

\bibitem{ZJ87}
O. Zeman and N-O. Jensen, Modification of turbulence characteristics
in flow over hills, Quart.  J. R. Met.  Soc.  {\bf 113}, 55-80(1987).

\bibitem{JH75}
P.S. Jackson and J.C.R. Hunt, Turbulent wind flow over a low hill,
Quart.  J. R. Met.  Soc.  {\bf 101}, 929-955 (1975).

\bibitem{HMGP78}
A.D. Howard, J.B. Morton, M. Gad el Hak and D.B. Pierce, Sand
transport model of barchan dune equilibrium, Sedimentology
{\bf 25}, 307-338 (1978).

\bibitem{JZ85}
N-O. Jensen and O. Zeman, in {\it International workshop on the
physics of blown sand}, Barndorff-Nielsen, Moller, Rasmussen and
Willets eds.  University of Aarhus, 351-368 (1985).

\bibitem{WG86}
F.K. Wippermann and G. Gross, The wind-induced shaping and migration
of an isolated dune: numerical experiment, Boundary-Layer Meteorology
{\bf 36}, 319-334 (1986).

\bibitem{HLR88}
J.C.R. Hunt, S. Leibovitch and K.J. Richards, Turbulent shear flows over
low hills, Quart.  J. R. Met.  Soc.  {\bf 114}, 1435-1470 (1988).

\bibitem{WHCWWLC91}
W.S. Weng, J.C.R. Hunt, D.J. Carruthers, A. Warren, G.F.S. Wiggs, I.
Livingstone and I. Castro, Air flow and sand transport over
sand-dunes, Acta Mechanica [Suppl] {\bf 2}, 1-22 (1991).

\bibitem{W95}
B.T. Werner, Aeolian dunes: computer simulations and attractor
interpretation, Geology {\bf 23}, 1107-1110 (1995).

\bibitem{BAD99}
J.H. van Boxel, S.M. Arens, P.M. van Dijk, Aeolian processes across
transverse dunes.  I: Modelling the air flow, Earth Surface Processes
and Landforms {\bf 24}, 255-270 (1999).

\bibitem{KSH01}
K. Kroy, G. Sauermann and H.J. Herrmann, A minimal model for sand
dunes, Phys.  Rev.  Lett.  \textbf{88}, 054301 (2002)

\bibitem{S01}
G. Sauermann, PhD Thesis, Stuttgart University, edited by Logos Verlag
(Berlin) (2001).

\bibitem{ACD02a}
B. Andreotti, P. Claudin and S. Douady, Selection of dune shapes and
velocities.  Part 1: Dynamics of sand, wind and barchans, Eur.  Phys.
J. B \textbf{28}, 321-339 (2002).

\bibitem{LL69}
K. Lettau and H.H. Lettau, Bulk transport of sand by the barchans of
La Pampa La Hoja in southern Peru, Zeitschrift f\"ur Geomorphologie
{\bf 13}, 182-195 (1969).

\bibitem{LSHK02}
A.R. Lima, G. Sauermann, H.J. Herrmann, K. Kroy, Modelling a dune
field, Physica A {\bf 310}, 487-500 (2002).

\bibitem{H03}
P. Hersen, On the crescentic shape of barchan dune,
submitted to EPJ. B. (2003).

\end{thebibliography}
\end{document}